# Electron beam precession for a serial crystallography experiment in a TEM

**Sergi Plana-Ruiz,[a,b]\* Penghan Lu,[c] Govind Ummethala,[c] Rafal E. Dunin-Borkowski.[c,d]**

[a] Servei de Recursos Científics i Tècnics, Universitat Rovira i Virgili, Avinguda Països Catalans 26, 43007 Tarragona, Catalonia, Spain.

[b] LENS-MIND, Department of Electronics and Biomedical Engineering, Universitat de Barcelona, Martí i Franquès 1, 08028 Barcelona, Catalonia, Spain.

[c] Ernst Ruska-Centre for Microscopy and Spectroscopy with Electrons and Peter Grünberg Institute, Forschungszentrum Jülich, Wilhelm-Johnen-Strasse, 52425 Jülich, Germany.

[d] Aachen University, Ahornstraße 55, 52074 Aachen, Germany.

Correspondence e-mail: sergi.plana@urv.cat

**Synopsis** The use of electron beam precession in a (S)TEM for serial crystallography experiments is thoroughly investigated, showing its potential for better crystal structure determination and refinement in the context of 3D electron diffraction.

**Abstract** During the last few years, serial electron crystallography (Serial Electron Diffraction, SerialED) has been gaining attention for the structure determination of crystalline compounds that are sensitive to the irradiation of the electron beam. By recording a single electron diffraction pattern per crystal, indexing hundreds or even thousands of measured particles, and merging the reflection intensities of the successfully indexed patterns, one can retrieve crystal structure models with strongly mitigated beam damage contributions. However, one of the technique's bottlenecks is the need to collect that many diffraction patterns, which, done in an automated way, results in low indexing rates. This work demonstrates how to overcome this limitation by performing the serial crystallography experiment following a semi-automated routine with a precessed electron beam (Serial Precession Electron Diffraction, SerialPED). The precession movement increases the number of reflections present in the diffraction patterns and dynamical effects related to specific orientations of the crystals with respect to the electron beam are smoothed out. This leads to more uniform reflection intensities across the serial dataset and a smaller number of patterns are required to merge the reflection intensities for good statistics. Furthermore, structure refinements based on the dynamical diffraction theory become accessible, providing a novel approach for more accurate structure models. In this context, the use of beam precession is presented as an advantageous tool for serial electron crystallography as it enables reliable crystal structure analysis with a lower amount of diffraction data.

## 1. Introduction

The field of serial crystallography aims at studying crystal structures by a collection of diffraction patterns from which each one corresponds to a randomly oriented single individual particle. This methodology was primarily developed in X-ray free electron lasers (XFELs) as a novel tool to study submicrometre-sized macromolecular crystals at the highest resolutions in space and time, which was one of the hindrances of biomolecular imaging at earlier times (Neutze *et al.*, 2000; Chapman *et al.*, 2011; Pellegrini, 2012). The use of a very bright X-ray beam pulsed at the femtosecond scale enables the illumination (and disintegration) of particles that are injected into the X-ray optical path through a constant stream, capturing the diffracting signal produced by each hit (Spence, 2017). The analysis of the resulting thousands to hundreds of thousands of effective diffraction patterns allows the elucidation of structures from crystals too small to be revealed by more conventional methods (Colletier *et al.*, 2016), as well as the dynamics of protein nanocrystals (Nass *et al.*, 2020). However, XFELs are not readily available to most labs, and high crystal densities are required per sample. In this context, the serial crystallography experiment in a transmission electron microscope (TEM) appears as an alternative and suggestive solution because electrons can be focused down to the Angstrom resolution, and their interaction with matter is stronger than X-rays or neutrons, which leads to a smaller number of quanta per elastic scattering interaction (Henderson, 1995; Clabbers & Abrahams, 2018).

Serial electron crystallography deals with electron diffraction (ED) patterns that do not have an *a priori* geometric relation between them. The first realization was done by using the TEM operation mode of the microscope (Smeets *et al.*, 2018), and later it was extended to the STEM mode (Bücker *et al.*, 2020). Essentially, the data acquisition workflow is the same iterative routine; an acquisition of a (S)TEM reference image for possible crystalline targets, automated/manual selection of electron beam positions, collection of a single/frame fractionated ED pattern for each chosen point, and shift of the stage to another interesting area. By following this protocol, the whole electro-transparent area of a typical TEM grid can be inspected, and thousands of patterns can be collected in a reasonable amount of time. Afterwards, the data processing takes place through self-made or adapted versions of pipelines from XFELs (Bücker *et al.*, 2021) that include: finding of central/primary beam and reflection positions (peaks) for each pattern, indexing with usually known unit cell parameters (some options exist for unknown cells (Jiang *et al.*, 2011; Gevorkov *et al.*, 2019)), reflection intensity extraction from the successfully indexed patterns, and merging of the extracted intensities.

One of the requisites for serial crystallography is the high number of diffraction patterns. The same requirements apply to electrons, where several hundreds to tens of thousands have been reported in other works (Smeets *et al.*, 2018; Bücker *et al.*, 2020; Nikbin *et al.*, 2024). However, during the 90s, several studies were made where the use of a few zone-axis ED patterns (ED patterns oriented at high symmetry axes) was shown to be enough to determine crystal structures (Morniroli & Steeds, 1992; Nicolopoulos *et al.*, 1995; Dorset, 1996, 1997). One of the disadvantages of this methodology was the time-consuming orientation of the crystals and the consequent unavoidable illumination of crystals before any meaningful acquisition, which is very critical for beam sensitive specimens. Furthermore, the effects of dynamical diffraction required very thin samples, and even then, these were not fully diminished as they are enhanced when oriented in zone axis. In this context, the combination of zone-axis ED patterns and high-resolution TEM images helped to push the accuracy of structure models characterized by electrons (Weirich *et al.*, 1996, 2006), but one of the biggest steps was the acquisition of diffraction data by beam precession, also known as rocking illumination (Vincent & Midgley, 1994).

Precession electron diffraction (PED) was invented to average out the non-systematic dynamical effects, such as Kikuchi lines, double diffraction or diffuse scattering, which are highly dependent on the crystal orientation, and render diffraction patterns with *pseudo*-kinematical reflection intensities, *i.e.*, reflection intensities that resemble more the respective calculated ones by the kinematical theory of diffraction. Crystal structure analysis from zone-axis PED patterns made crystal structure analyses easier (Weirich *et al.*, 2006; Gemmi & Nicolopoulos, 2007; Sinkler *et al.*, 2007). Later on, the idea of three-dimensional electron diffraction (3D ED) was introduced (Kolb *et al.*, 2007, 2008; Gemmi *et al.*, 2019); the collection of non-oriented ED patterns from a single nanocrystal at subsequent and usually equidistant tilts of the sample holder. Here, the addition of precession also resulted in a significant improvement (Mugnaioli *et al.*, 2009) as the ranking of reflection intensities was better preserved for *ab initio* structure solutions (Klein & David, 2011; Eggeman & Midgley, 2012), and subsequently enabled dynamical refinements (Palatinus *et al.*, 2013). From another perspective, the addition of precession to scanning electron diffraction (SED), known as 4D-STEM in other literature (Ophus, 2019), also resulted in better results, for instance, for phase and orientation mapping (Viladot *et al.*, 2013). The enhanced quality of these maps comes from the wobbling of the Ewald sphere by beam precession since it swaps/integrates a larger volume in the diffraction space, which leads to more reflections per ED pattern with less dynamically related noise. In this way, indexing algorithms like template-matching work better (Rauch & Dupuy, 2005; Rauch *et al.*, 2010). Other SED applications like strain mapping (Cooper *et al.*, 2015) or electric field mapping (Lorenzen *et al.*, 2024) also benefit from precession, but in this case, the advantage is related to the uniformization of the intensity inside the reflection disks.

Given the history of success for PED, this work aims to evaluate the benefits of precession in the context of serial electron crystallography for crystal structure determination and refinement. The analysis described hereunder is performed from ED patterns acquired with and without precession from a beam stable sample at different microscope setups and processed with different software/algorithms. In this way, a detailed and fair comparison of the retrieved and refined structure models between static and precessed serial data is presented.

## 2. Materials and Methods

$BaSO_4$ (barite) was used for the comparisons between statically acquired (static) and precession-integrated (precessed) serial data, referenced in this work as SerialED and SerialPED, respectively. Barite is an inorganic material crystallizing in an orthorhombic space group (7.154 Å, 5.454 Å, 8.879 Å; *Pnma*) up to very high resolutions (more than 2 Å$^{-1}$) (Jacobsen *et al.*, 1998), and the electron irradiation does not diminish its crystalline state; hence its use as a reference in other ED works (Mugnaioli *et al.*, 2009; Plana-Ruiz *et al.*, 2020). Fine powder of this compound was purchased from Merck, dispersed in ethanol and casted onto typical Cu-TEM grids (ultra-thin continuous amorphous C).

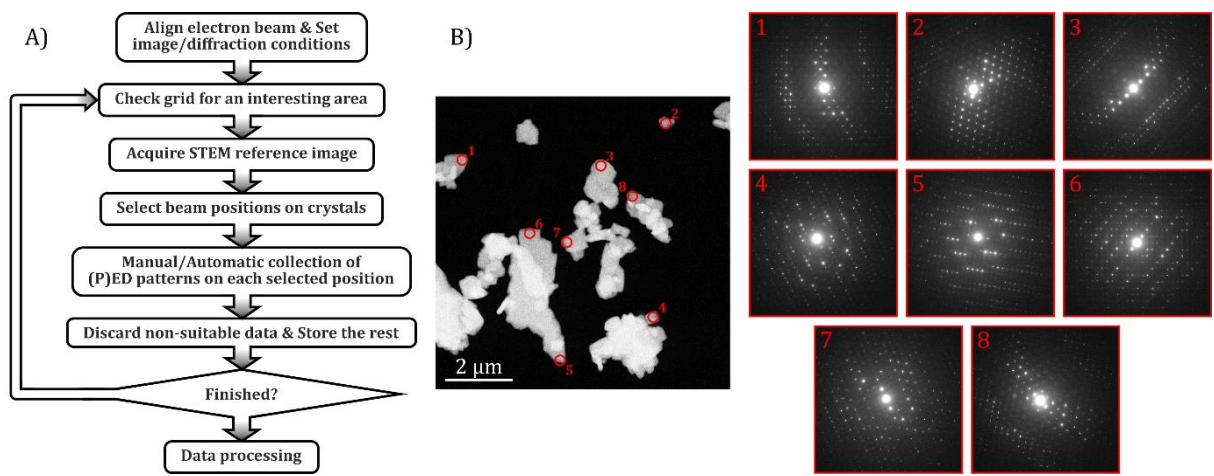

**Figure 1** A) Schematic of the semi-automated Serial(P)ED experiment workflow and B) an example of a STEM-HAADF reference image and the respective PED patterns acquired in each red marker.

The acquisition of the static and precessed SerialED data was performed with two different TEMs equipped with two different detectors. The first one is a JEOL F200 ColdFEG operated at 200 kV (0.02517 Å) in STEM mode (Probe size 8, 10-µm condenser aperture) with a post-column Gatan OneView camera (16-bit CMOS-based and optical fibre-coupled detector of 4096 x 4096 pixels; 15-µm physical pixel size). The STEM operation mode was chosen instead of the TEM mode because it is better suited for diffraction experiments, as already reported (Kolb *et al.*, 2019; Hogan-Lamarre *et al.*, 2024). A quasi-parallel beam in STEM was manually aligned following the routine established by (Plana-Ruiz *et al.*, 2018), so that the most parallel condition could be attained with a beam diameter of 200 nm (FWHM) for the diffraction pattern acquisition. A more convergent electron probe was used for the STEM reference images acquired with a JEOL high-angle annular dark-field (HAADF) detector. A Gatan Digital Micrograph script was developed to facilitate the data collections with a graphical user interface for its ease-of-use (Figure S1). Briefly, it allows one to save and retrieve beam conditions for imaging and diffraction settings (beam size, camera length and projector coils offsets), acquire STEM reference images, manually select as many beam positions as one wishes from this reference, and automatically shifts the beam, and collects and stores the ED patterns (freely available at *github.com/sergiPlana/TEMEDtools/tree/main/STEMSerialED*). The precession of the electron beam at 100 Hz frequency was enabled by a P2000 prototype precession unit provided by NanoMEGAS SPRL. Precession-assisted 3D ED tilt-series data was automatically collected using the Fast-ADT module with a JEOL tomography holder that allows a maximum tilt range of +/- 70° (Plana-Ruiz *et al.*, 2020). The second microscope used in this work was the TESCAN Tensor, a STEM-dedicated Schottky FEG operated at 100 kV (0.03701 Å) with a Dectris Quadro detector (16-bit hybrid-pixel direct electron detector of 512 x 512 pixels; 75-µm physical pixel size). The interface ExpertPI based on Python v11.6 was used for the rapid switch of the two different beam settings, the acquisition of STEM bright-field (BF) reference images, and the acquisition of precessed and static diffraction patterns from manually selected positions. A 200-nm beam-diameter was set for the collection of the patterns. Precession of the electron beam was enabled at a frequency of 72 kHz from the signal unit integrated into the microscope. Figure 1 shows the acquisition workflow followed in this study.

Different software packages were employed to process the diffraction data. Data reduction (from raw frames to reflection intensity / *hkl* files) was independently obtained from two pipelines for comparison purposes: *PETS2* v2.2.20240601 (Palatinus *et al.*, 2019) and the *diffractem* v0.4.0 Python package (Bücker *et al.*, 2021) that uses routines from the *CrystFEL* software suite v0.10.0 (White *et al.*, 2012). For the latter, the indexing was retrieved by the *pinkIndexer* algorithm (Gevorkov *et al.*, 2020) available from the *indexamajig* program of *CrystFEL*, and the merging of all integrated reflection intensities was done *via* the *partialator* program of *CrystFEL*, which includes scaling, different models for the calculation of partial intensities, and post-refinement (White, 2014). *Ab initio* structure solutions were obtained by direct methods in *Sir2014* v17.10 (Burla *et al.*, 2015) using the *BEA* algorithm for better results (Luca Cascarano *et al.*, 2010), and the charge-flipping algorithm in *SUPERFLIP* v09.21.20 (Palatinus & Chapuis, 2007). Crystal structure refinements were done with *Jana2020* v1.3.57 (Petříček *et al.*, 2023). Dynamical refinements were executed in *Jana2020* v1.3.57 using the *dyngo* module (Palatinus, Petříček *et al.*, 2015). Visualization of the structure models was obtained from VESTA3 (Momma & Izumi, 2011).

## 3. Results

200 barite crystals were measured with and without 0.92° of precession integration across 45 reference images with fields of view between 2.7 x 2.7 µm$^2$ and 8.7 x 8.7 µm$^2$ with the 200kV TEM for ∼ 1.5 hours. Additionally, 495 crystals were inspected with and without 0.97° of precession from 38 reference areas of 12.5 x 12.5 µm$^2$ with the 100kV microscope for ∼ 3 hours. Since the two sets of serial diffraction data come from different electron energies and electron optics, and the reflection intensities were also detected from two different technologies, their processing was done separately. Figure 2 shows representative reference images from both setups for crystal measurement selection.

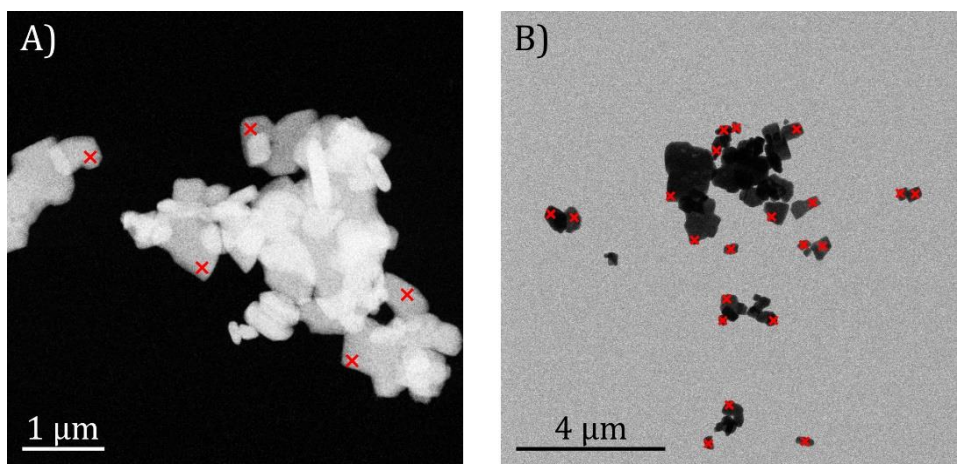

**Figure 2** Representative reference images used to select interesting crystalline $BaSO_4$ particles. The red crosses correspond to the positions where the beam was placed to acquire ED patterns. A) STEM-HAADF image from the JEOL F200 TEM at 200 kV and B) STEM-BF image from the TESCAN Tensor microscope at 100 kV.

Two software packages were used for data reduction to compare two different ways of reflection intensity extraction as well as of indexing procedures. The *PETS2* package offers the possibility to extract the reflection intensities based on the fitting of specific functions to the shape of the rocking curves. For indexing, a template-matching algorithm has been added recently that assigns alpha, beta and gamma orientation angles to each pattern as if it would be a 3D electron diffraction (3D ED) dataset and enables to process it likewise (Palatinus *et al.*, 2023). On the other hand, *diffractem*/*CrystFEL* offers the possibility to consider the partiality of reflection intensities from different geometric models for intensity extraction, i.e., that the integrated intensity from a given reflection does not come from the Bragg condition (White, 2014). From the indexing perspective, the algorithm of *pinkIndexer* is presented as an alternative by parametrizing the possible orientations of the crystal lattice as curves in a 3D rotation space, which has been successfully tested on X-ray and electron diffraction (Gevorkov *et al.*, 2020; Bücker *et al.*, 2020; Hogan-Lamarre *et al.*, 2024). Table 1 shows some of the statistics of the resulting data reduction process from the Serial(P)ED experiments using these two processing pipelines.

**Table 1** Data reduction statistics for the Serial(P)ED data collected from barite crystals on the different microscope setups. Data processing according to "profile fit" was obtained from *PETS2*, while "scaling refinement" was done through *diffractem*/*CrystFEL* with three post-refinement iterations. "Used patterns" refers to the number of patterns that have been correctly indexed. "Refls" stands for reflections, and "Ind." for independent. "Integrated Refls" represents the total number of integrated reflections through the whole serial dataset without merging. Reflections up to 2 Å$^{-1}$ resolution have been considered.

| Electron energy (keV) | 200 | | | | 100 | |
|---|---|---|---|---|---|---|
| Refl. intensity extraction | **Profile fit** | | **Scaling refinement** | | **Profile fit** | |
| Data acquisition approach | **Static** | **Precessed** | **Static** | **Precessed** | **Static** | **Precessed** |
| Used patterns (#) | 174 | 199 | 193 | 189 | 472 | 485 |
| Percentage of all patterns (%) | 87.0 | 99.5 | 96.5 | 94.5 | 95.4 | 98.0 |
| Integrated Refls (#) | 20035 | 53264 | 20706 | 22719 | 28310 | 129939 |
| Merged Refls (#) | 3067 | 8258 | 8838 | 9437 | 3970 | 9227 |
| Ind. Refls* (#) | 1083 | 1478 | 1320 | 1364 | 960 | 1477 |
| Completeness* (%) | 70.01 | 95.54 | 85.33 | 88.17 | 62.14 | 99.06 |
| $R_{int}$* (%) | 31.96 | 11.60 | 44.44 | 16.76 | 23.31 | 13.07 |

* As calculated by *Sir2014* for reflections that fulfil that their intensity is above 3σ(I).

An important point for the data reduction is to determine if the found crystal orientations (indexing) are correct. When dealing with thousands or even millions of diffraction patterns, filters are available to discard the incorrectly indexed patterns that are most obvious, and the small fraction that goes through as correct does not tend to have a high contribution since the overall averaging smears them. However, if the number of patterns is small, incorrectly indexed patterns should be excluded to avoid any significant biasing. In this work, the criterion was set to be when the resulting indexing (if given) provided meaningful reflection positions with respect to the experimental ones by visual inspection. In the case of *PETS2*, the refined frame scales obtained from each pattern were also checked, discarding the ones that were negative or exceedingly high. In this way, the rate of used patterns after parameter optimizations and safety checks was higher than 85% in all evaluated cases (See Table 1).

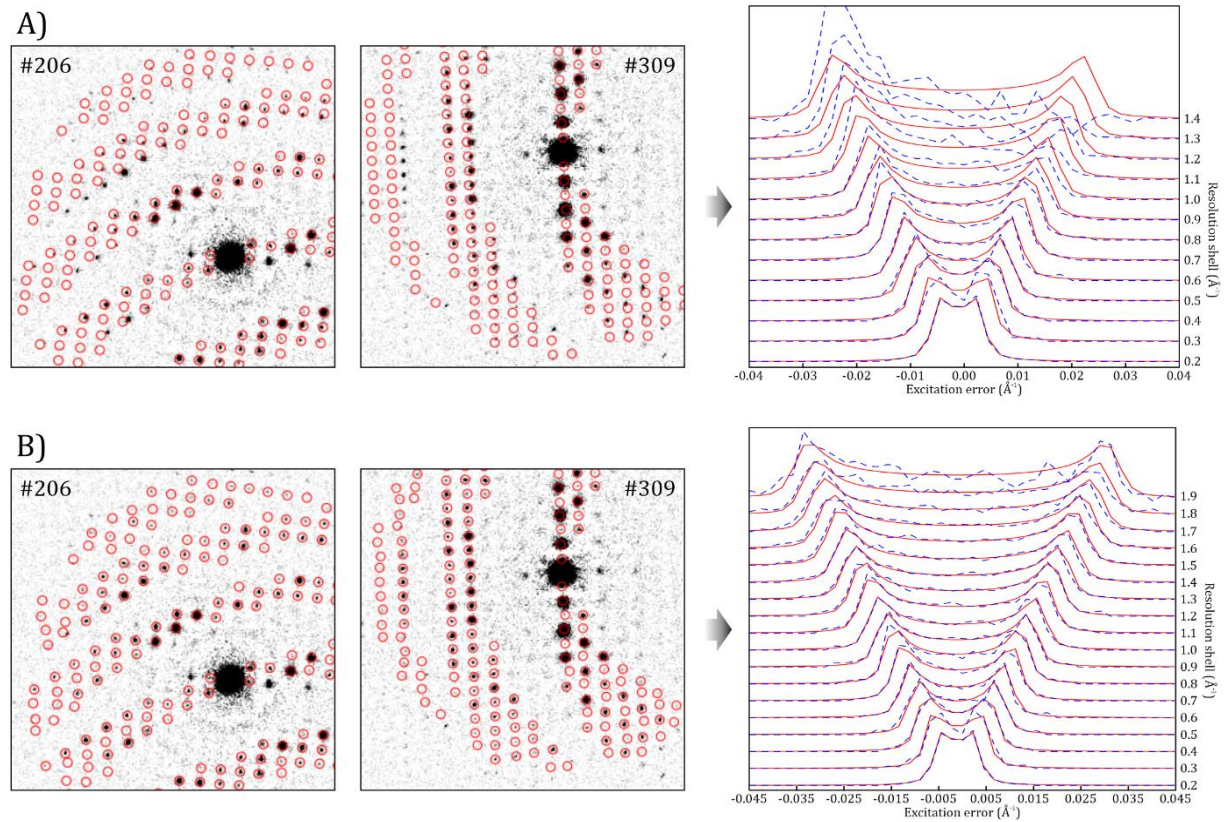

**Figure 3** Two exemplifying background-corrected diffraction patterns of the SerialPED experiment from the 100kV microscope with overlayed red circles that correspond to the calculated reflection positions according to the found orientation, and the resulting averaged rocking curves at different resolution shells considering all diffraction patterns from *PETS2*. Blue-dashed curves represent the averaged experimental result, and the red ones the simulated double-peaked curve that fits best. A) case in which reflection positions are treated as completely free of distortions and B) when distortion corrections are enabled.

During the data reduction process, it was noticed that the diffraction patterns acquired at 100kV exhibited a strong distortion, which was only perceptible when the calculated reflection positions according to the orientation matrix were overlapped with the experimental patterns, and the averaged rocking curves at different resolution shells were plotted for the precessed data. Figure 3A displays two PED patterns from where this mismatch can be clearly seen and the strong asymmetry of the double-peaked rocking curves when processing the whole respective dataset. This results in very poor intensity integration for reflections far away from the central beam that leads to worse intensity statistics as resolution increases, poor least-squares fitting of the function parameters to the experimental averaged rocking curves, and thus incorrect reflection intensity extraction at the end. Unfortunately, this was not only an elliptical distortion caused by residual stigmatism of the projector system, but a combination of several distortions of higher order that is suspected to be due to the data acquisition at an optical plane not exactly conjugated to the back focal plane of the objective lens. In this situation, the optical distortions were corrected by applying the available option in *PETS2* (Brázda *et al.*, 2022); first, the frame-by-frame distortions that include magnification, elliptical and parabolic correcting factors were obtained by least-squares fitting on each pattern, and then the barrel-pincushion was determined by least-squares refinement on the 3D reconstruction of the 2D peak positions (the 'radial $S_g$ parabolic' parameter related to the dependence of the parabolic distortion with the phase of the precession circuit was also refined in case of precessed data). These last two contributions were the dominant as they reached -0.92% for static patterns, and -0.85% and -0.71%, respectively, for the precessed ones. The result of this detailed geometrical correction is shown in Figure 3B, where the calculated positions of the reflections match well the experimental ones and the double-peaked rocking curves appear symmetric up to very high resolutions. It is important to note that such corrections are applied to the reflection positions from which the respective intensity will be integrated. Thus, no image transformation is applied to the frames, and the reconstructed observable diffraction space will still exhibit the deformation (See Figure S2). On the other hand, the *diffractem* package has only the option to correct for the typical elliptical deformation, hence equivalent data processing comparison could not be made for the serial dataset at 100 kV.

As mentioned above, the use of *partialator* includes several options to merge the reflection intensities that have been previously reported to not significantly influence the final outcome (Bücker *et al.*, 2021). However, the analysis carried out here shows that the choice of these parameters for the SerialED dataset determines if a successful structure solution is possible (understood as finding the maximum number of atoms), although the figures-of-merit (FoM) are not good, such as negative overall atomic displacement parameters (ADPs) and high $R_{int}$s (See Table S1). The case of the precessed dataset is totally different as the merging is more uniform across the routines landscape (See Table S2); all used merging options resulted in at least 4 of the 5 symmetry-independent atoms, the $R_{int}$s were significantly better compared to the static case, and the overall ADPs were positive. Kinematical refinements followed for both types of serial data by using the merged *hkl* file that resulted in the best performance; unity model (partialities = 1 for all reflections), Debye-Waller scaling calculation, and 3 iterations of post-refinement.

The kinematical refinements were carried out in *Jana2020* using the structure models obtained with charge-flipping, which did not appreciably change with respect to the ones obtained from direct methods. Table 2 and Table S3 show the FoM for the SerialPED and SerialED results, respectively. The structures could only be refined with isotropic ADPs, and they became positive in all the precessed cases considered here, while some of them turned negative for the static patterns. Interestingly, all analysis showed that the atom that had more problems to be refined is the O3, which always tended to be too close to the Sulphur (below 1.3 Å), and the isotropic ADP was higher or even non-refinable. Furthermore, FoM are also quite high for electron diffraction standards.

Since kinematical refinements led to distorted tetrahedron with too small S-O3 distances, dynamical refinements from the precessed diffraction patterns were performed to improve the crystal structure model and the overall FoM. Usually, such refinement can only be performed on 3D ED datasets as the reflection intensities in the dynamical theory of diffraction are strongly dependent on the thickness of the crystal, and a single thickness value is refined for the tilt series of the individual crystal. Each pattern and its related integrated reflection intensities are treated individually, considering its crystallographic orientation and the increase of the virtual thickness due to the alpha-tilt of the sample holder (as the tilt increases, the distance that the electrons go through the crystal increases as well). The dynamical refinement module in *Jana2020* has an option to automatically apply this thickness correction that can be switched off. On the other hand, *PETS2* has the possibility to apply the frame scales found for each diffraction pattern (used for the creation of the *hkl* file and based on the Laue class symmetry) to the integrated reflection intensities of each pattern. In this way, the reflection intensities across the different patterns are comparable and the dependence on the thickness could be smoothed, assigning a single virtual thickness value for all the SerialPED dataset to be refined through the dynamical calculation procedure. Although it is not formerly correct, the ideal situation to refine the thickness for each frame leads to unstable refinements, which leaves this consideration as a heuristic approximation for better crystal structure refinements of serial electron diffraction data. The FoM for the dynamical refinements from both sets of SerialPED datasets are shown in Table 2. From both SerialPED datasets, FoM became significantly better, the geometry for the Sulphur-Oxygen chemical environment resembles an ideal tetrahedron, and the root mean square deviation (RMSD) was reduced to the picometer scale in comparison to the kinematically-refined structures.

**Table 2** Figures-of-merit for the structure refinements carried out in *Jana2020* for the SerialPED data collected from barite crystals on the different microscope setups. "Profile fit Int." stands for profile fit intensity extraction of the *hkl* file obtained from *PETS2*, while "Scaling ref. Int" stands for scaling refinement intensity extraction of the *hkl* file from *diffractem*/*CrystFEL* using three post-refinement iterations. RMSD correspond to the root mean square deviation of the atom positions as calculated by *Sir2014*, and "Max. deviation" refers to the maximum distance variation of the atom positions with respect to the structure model from (Jacobsen *et al.*, 1998) as reference.

| Electron energy (keV) | 200 | | | 100 | |
|---|---|---|---|---|---|
| Refinement type | **Kinematical** (Scaling ref. Int.) | **Kinematical** (Profile fit Int.) | **Dynamical** | **Kinematical** | **Dynamical** |
| Number of reflections (#) | 1350/1364 | 1194/1471 | 19257/20891 | 1190/1477 | 16602/53534 |
| Reflections/Parameters (-) | 79.4 | 70.2 | 83 | 70.0 | 32.1 |
| GoF (%) | 35.07/34.89 | 6.19/5.65 | 7.77/7.47 | 8.04/7.27 | 2.52/1.56 |
| $R$ (%) | 33.13/33.26 | 28.66/30.12 | 14.77/15.23 | 30.05/32.13 | 11.07/20.71 |
| $R_w$ (%) | 43.61/43.61 | 36.22/36.50 | 17.01/17.04 | 38.29/38.46 | 11.70/12.81 |
| RMSD (Å) | 0.072 | 0.047 | 0.008 | 0.076 | 0.010 |
| Max. deviation (Å) | 0.34(3) | 0.21(3) | 0.022(4) | 0.23(10) | 0.035(3) |

The number of reflections, goodness of fit (*GoF*), $R$ and $R_w$ parameters are calculated and reported from observed and all (obs/all) reflections up to 2 Å$^{-1}$ resolution. The criterion for observed reflections was $I(\mathbf{g}) > 3\sigma(\mathbf{g})$. The 'Reflections/Parameters' ratio refers to the number of observed reflections over the number of refined parameters. $R$ and $R_w$ are based on the square root of reflection intensities. Dynamical refinements were executed with $g_{max}$ of 2.2 Å$^{-1}$, $S_g^{max}$(matrix) of 0.01 Å$^{-1}$, $S_g^{max}$(refine) of 0.1 Å$^{-1}$, $RS_g$ of 0.66, and $N_{or}$ of 83 for the 200keV data and 87 for the 100keV one.

As a summary, Figure 4 shows the different methods used for the processing of the SerialED and SerialPED data according to the electron energy/microscope, and the respective type of crystal structure refinement performed for each extracted *hkl* file.

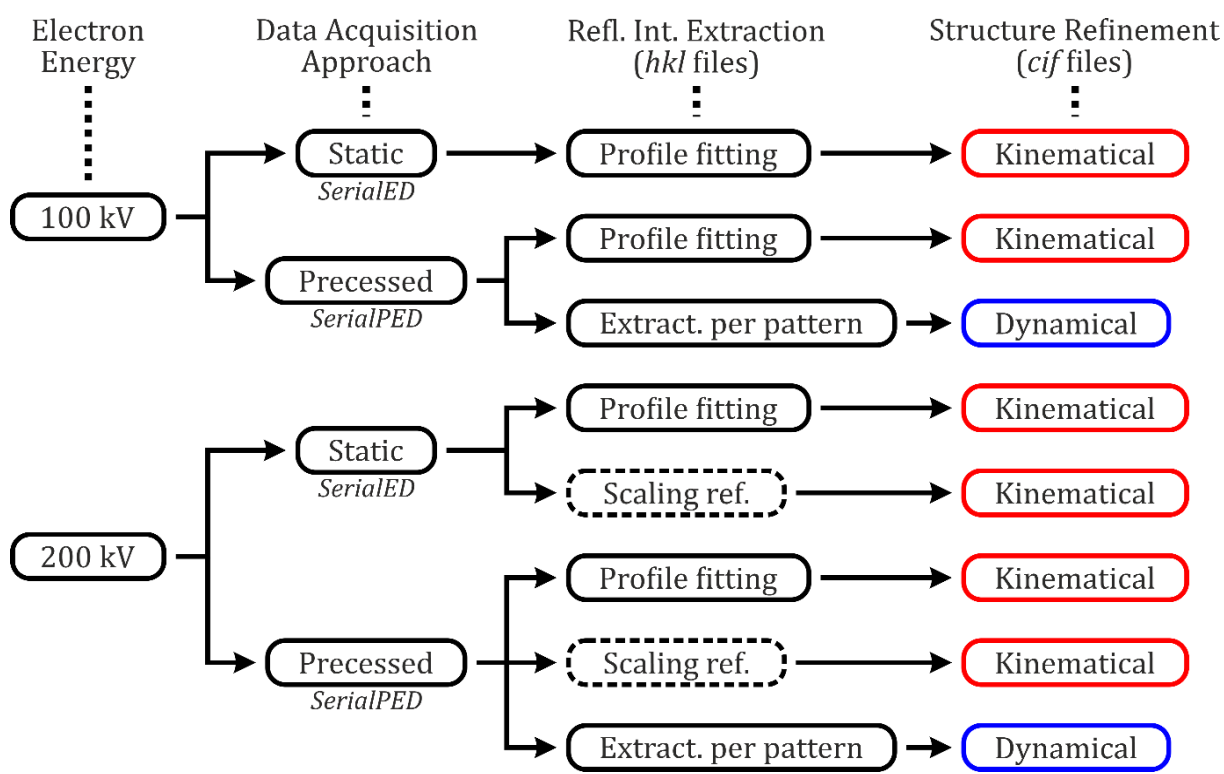

**Figure 4** Scheme summarizing the different data processing procedures for the extraction of the reflection intensities (*hkl* files) with respect to the collected serial data, and the type of structure refinement carried out in each case (*cif* files as final output): kinematical highlighted in red and dynamical in blue. "Profile fitting" and "Extract. per pattern" reflection intensity methods were performed in *PETS2*, and "Scaling ref." (surrounded by black dashed lines) in *diffractem/CrystFEL*.

## 4. Discussion

The case study presented here aims to assess if precession helps to improve the quality of SerialED data and thus, the crystal structure determination and refinement thereafter. Previous works focused on highly symmetric structures where the aim was to get enough ED data before the crystalline integrity of the material was vanished (Bücker *et al.*, 2020), and demonstrate the higher resolution that one can achieve in comparison to tilt-series experiments (Hogan-Lamarre *et al.*, 2024). In this context, the analysis of a lower symmetry inorganic crystal like $BaSO_4$ becomes relevant because it allows to explore the scenario of serial crystallography with a low number of patterns, hence less symmetrically related reflections that will be merged together. The results show that this is critical in the case of SerialED data, where the $R_{int}$ becomes significantly better for the set of 495 pattern (~ 23%) than the 200 frames' dataset (~ 41%) independently of the used merging protocol. Therefore, the idea that merging diffraction data from different crystals allows to smooth the dynamical effects, which does not only include reflection intensity re-distribution by multiple scattering but also Kikuchi lines contributions, and enables a list of *pseudo*-kinematic reflections becomes directly apparent. On the other hand, the use of precession achieves this same situation in each individual pattern, reducing the required number of crystals to be measured for reliable structure analysis. See Figure 5 for the strong influence of dynamic effects on the reflections that is smoothed out when precession is applied to the same crystal.

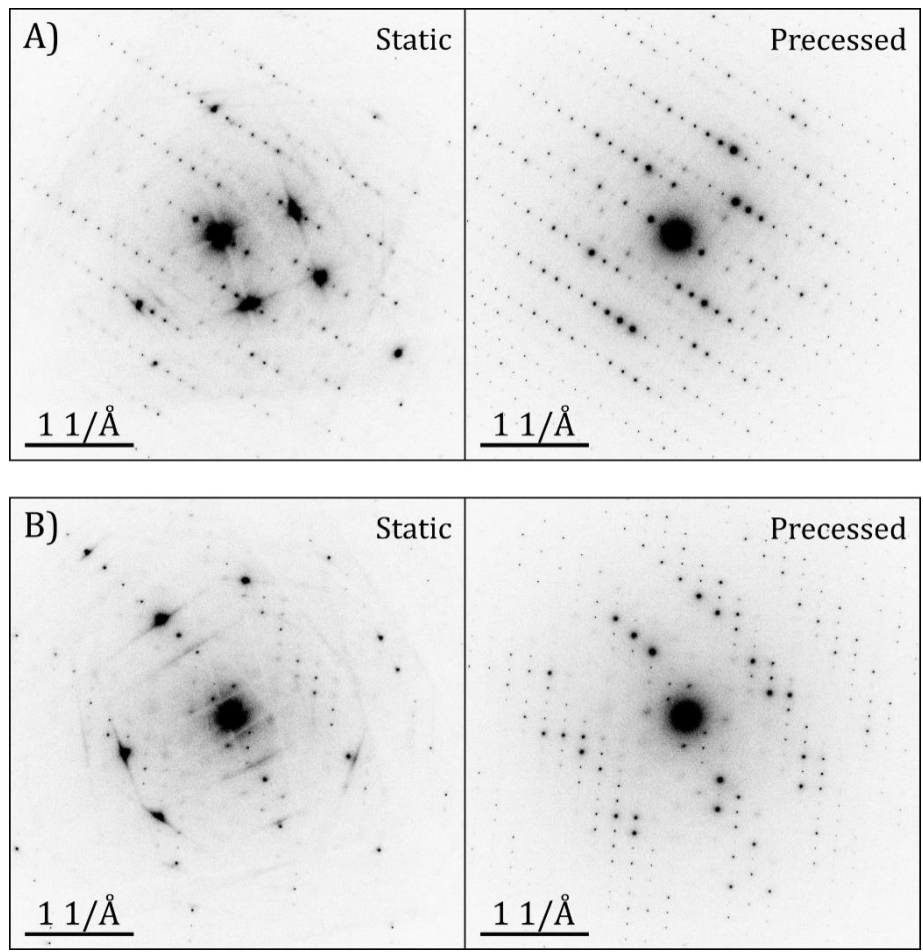

**Figure 5** Two pairs of ED patterns without (static) and with 0.92° of precession (precessed) from barite crystals where the effect of precession on the quality of the reflection intensities is directly visible. The displayed contrast on all patterns is the same for the most suitable comparison.

One of the advantages of integrating a volume from the observable diffraction space into an ED pattern by beam precession is the possibility to perform much more accurate refinements using the dynamical theory of diffraction. Minimizing or smearing the dynamical effects does not mean that reflection intensities are not intrinsically dynamical anymore, and the procedure followed here for the dynamical refinements demonstrates how the models become more accurate and reliable. The geometry similarity between the reference X-ray model and the found ED structures is much better from the dynamical refinement than for the kinematical case (RMSD reduced by an order of magnitude), and the ADPs could be anisotropically refined resulting in positive values for all diagonal elements. The kinematical refinements could only be performed with isotropic ADPs and some of them still become negative. Finally, the overall FoM become significantly better, which the $R$ and $R_w$ figures decreased by more than a factor of 2; from ∼ 31% to ∼ 13% for $R$, and from ∼ 39% to ∼ 14% for $R_w$ (on average for observed reflections).

Preferred orientation was spotted on the reconstructed $0kl$ and $h0l$ sections of the observable diffraction space (See Figure S2), but the completeness was high enough to avoid significant missing wedge effects on the retrieved electrostatic potential. Nevertheless, elongation of the anisotropic ADPs was observed along the $c$-direction corresponding to the main direction of the missing wedge. To discard the possibility that such effect is a result of any inappropriate data processing step, tilt-series diffraction data was collected on two different crystals from the same grid used for the serial acquisition, and crystal structure determinations and dynamical refinements followed using the usual procedure (See Table S4). Figure 6 shows the refined structure models along $b$ for comparison. In all cases, the trend of the ADPs geometry is very similar. Interestingly, the strong anisotropy obtained from the serial data is also visualized on the model of Figure 6C. The inspection of the diffraction space confirmed that the missing wedge is comparable in both cases, hence the similarity in the resulting crystal structures (See Figure S3). On the other hand, the diffraction space of Figure 6B model has a reduced missing wedge due to the high angular range, which explains the closer isotropy of the refined ADPs. This comparison between 3D ED acquisition techniques demonstrates the validity of the presented dynamical refinement approach on serial data.

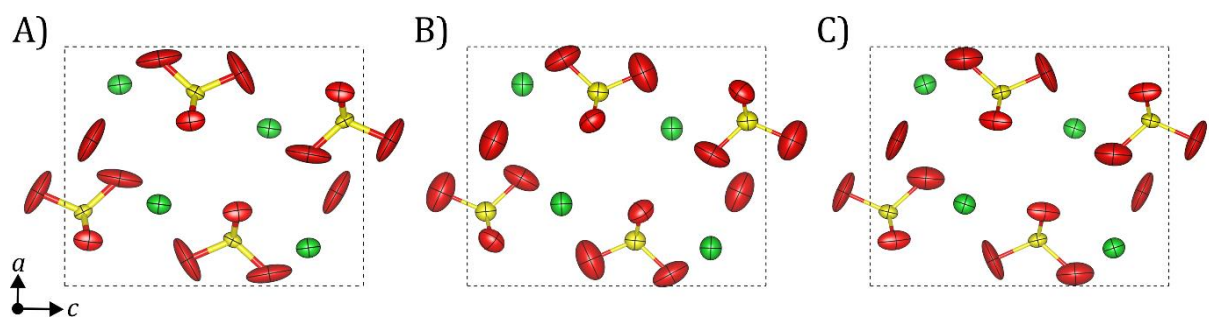

**Figure 6** Dynamically refined structure models of barite projected along *b* from precessed A) serial and B)-C) tilt-series 3D ED data. Diffraction data collected at the 200kV microscope on the same TEM grid, same illumination conditions and same detector parameters. Model from B) was obtained from 121 diffraction patterns expanding 120° of angular range, while C) corresponds to 101 patterns across 100°.

Although it is interesting to consider the different parameters for the merging of reflection intensities in the context of kinematical diffraction, this work has found that the fluctuating intensities given by the dynamical nature of electron diffraction has a strong effect (See Table S1). Furthermore, the case of a low number of patterns and relatively low symmetry implies that most reflections may not be detected more than once, and for static ED patterns, the geometric model and post-refinement merging iterations play a role in finding the best way to merge them. If the frame scales per pattern obtained by *PETS2* are compared between static and precessed diffraction datasets, the difference between the (P)ED patterns can be qualitatively quantified, which demonstrates the suitability of precession in these situations. For SerialED, the mean frame scales were 0.92 ± 0.68 and 0.98 ± 0.63 for 100kV and 200kV, respectively. In SerialPED, they become 1.00 ± 0.54 for the former and 1.00 ± 0.28 for the latter. The standard deviation is smaller in both electron energy cases, indicating the presence of more uniform reflection intensities across the ED patterns acquired with precession. This can also be visualized on the averaged rocking curves, which become noisier, and the profile fitting is worse for the static collection (See Figure S4).

The use of different intensity extraction and indexing algorithms allowed the evaluation of different data processing pipelines. The main significant difference is found for static data. Here, the profile fit resulted in a lower number of successfully indexed patterns and a lower number of merged reflections (See Table 1). Completeness is thus lower as well, but the final $R_{int}$ is better compared to the scaling refinement of *partialator* (See Table 2). The FoM are also better for the profile fit, but the respectively refined structure converged with Ba, S and one O with negative ADPs. The model from *partialator* reflections leads to very low ADPs and the one from Ba became negative. The kinematical refinement also shows that the number of reflections above 3σ is very low for the profile fit, which could be explained by the high dynamical effects present in the ED data that cannot be compensated by the frame scales, hence high standard uncertainties are assigned during the fitting of the reflection rocking curves. From the precession point of view, *PETS2* considers the geometry of precession to calculate the theoretical positions of the reflection intensities, which results in more integrated reflections per pattern compared to the static case. *indexamajig* does not consider it and the only way to induce the software to integrate more reflections is to increase the reflection profile radius, yet it is still far away from the number of reflections contemplated by a precession experiment. However, the kinematically refined structures are very similar in terms of FoM and ADPs (See Table 2). The indexing algorithms did not perform differently either, the only key aspect is the correct pixel calibration, which strongly determines the successful indexing, and both pipelines incorporate tools to optimize and/or refine parameters to increase the respective score functions.

Finally, one of the strong aspects of *PETS2* is the correction of optical distortions on the ED data (Brázda *et al.*, 2022). This is an important step as it is very common to acquire ED patterns at an optical plane that is not exactly conjugated to the back-focal plane of the objective lens, for instance, by not using the standard lens currents of the objective and diffraction lenses. It becomes crucial for the correct reflection integration as the resolution increases, which is certainly key for the structure determination of complex compounds. One may wonder if the strongly observed distortions could affect the focusing of the reflections when precession is used, *i.e.*, whether reflection splitting could be problematic. If one does not consider applying non-linear offsets at the different phases of the sinusoidal signals of precession (Viladot *et al.*, 2013), it comes down to the number of pixels and the point-spread function of the detector as well as the effective camera length. This means that given the same diffractive object and the same effective camera length (equal resolution at the edge of the detector), the intensity counts for a given reflection will be spread around more pixels on detectors with a higher number of pixels. Thus, the possibility of observing splitting is higher for these ones. In this work, the pixel calibration for the optical fibre-coupled detector was 0.001 $Å^{-1}$, and for the direct detector was 0.011 $Å^{-1}$, extending the intensity for a given reflection on a circular area of around 50 pixels in diameter for the former, and 5-6 pixels for the latter. Therefore, if splitting occurs with a maximum deviation of 0.01 $Å^{-1}$, it will be seen with the indirect detector as it will represent an elongation of the reflection intensity of ~ 10 pixels in a specific direction, while it will not be observed with the direct detector as all electrons with such angular spread will fall on the same pixel/cell (for simplicity, the point-spread function has not been considered). Nonetheless, a careful optical alignment of the electron beam should always be carried out to avoid such distortions as much as possible and get the most from the diffraction space.

**5. Conclusions**

The thorough investigation carried out here from the ED datasets of $BaSO_4$ shows that the use of precession in a serial electron crystallography experiment is advantageous. The presented work demonstrates that PED helps to extract *pseudo*-kinematic reflection files with fewer diffraction patterns. The requisite of a high number of patterns in a typical SerialED acquisition is mandated primarily by the smoothing of the dynamical effects through the averaging across the serial dataset, but the use of precession accomplishes this smearing in each pattern, hence reducing the number of crystals to be measured. In the past, such a line of thought was applied to zone-axis patterns, where precession helped to diminish the dynamical effects, but it was not always enough to have a successful structure determination from the merging of a few patterns following the kinematical approach (Gjønnes *et al.*, 1998). Combining a few hundreds of randomly oriented crystals with precession seems to be the ideal experimental setup for crystal structure analysis. Furthermore, dynamical refinements become accessible due to the integration of a small fraction of the diffraction space into each individual pattern, which can be performed to increase the accuracy of the structure models by a factor of magnitude, similar to what was observed with tilt-series 3D ED data (Palatinus, Corrêa *et al.*, 2015; Klar *et al.*, 2023).

The lowering of the required amount of diffraction patterns for successful and accurate structure determinations opens the possibility to use such data acquisition methodology in other laboratories that do not have automated collection protocols. This type of data can be acquired in a reasonable amount of time following a semi-automated way. Moreover, the stage does not have to be optimized for tomography (tilt/rotation) experiments, expanding the range of microscopes that can be utilized for this purpose. These findings will be of special benefit to the investigation of beam-sensitive materials such as metal-organic frameworks, molecular crystals or even macromolecules that cannot be synthesized in very large amounts and stand very low electron fluences ($e^-/Å^2$), providing a reliable alternative approach for their structural analysis based on electron diffraction.

**Acknowledgements** The authors would like to thank the support from the TESCAN company during the measurements with the Tensor microscope.

**Conflicts of interest** There are no conflicts of interest.

# Supporting information

**Table S1** Merging statistics and results from *ab initio* structure solutions (*Sir2014*) of the SerialED data collected from barite crystals on the 200kV TEM according to the different options of *partialator* in the *CrystFEL* software package. "PR ite." stands for post-refinement iterations, "Ind. Refls" for symmetrically independent reflections, and "Compl." for completeness. Overall ADP is the isotropic atomic displacement parameter obtained from the Wilson plot and given as B. Reflections up to 2 Å$^{-1}$ resolution have been considered.

| PR ite. | Partiality model | Debye Waller calculation | Ind. Refls* (#) | $R_{int}$*(%) | Compl.* (%) | Overall ADP* (Å$^2$) | Found atoms from direct methods |
|---|---|---|---|---|---|---|---|
| 0 | Offset | No | 1220 | 35.11 | 78.86 | -0.090 | Ba, S, O1, O2 |
| 0 | Offset | Yes | 1230 | 36.04 | 79.51 | -0.069 | Ba, S, O1, O2 |
| 0 | Unity | No | 1325 | 44.00 | 85.65 | -0.104 | Ba, S, O1, O2, O3 |
| 0 | Unity | Yes | 1342 | 44.55 | 86.75 | -0.071 | Ba, S, O1, O2 |
| 0 | Xsphere | No | 1193 | 35.60 | 77.17 | -0.225 | Ba, S |
| 0 | Xsphere | Yes | 1194 | 37.01 | 77.23 | -0.392 | Ba, S, O1, O2 |
| 1 | Offset | No | 1221 | 34.03 | 78.93 | -0.106 | Ba, S, O1, O2 |
| 1 | Offset | Yes | 1233 | 34.94 | 79.9 | -0.67 | Ba, S, O1 |
| 1 | Unity | No | 1323 | 44.16 | 85.52 | -0.113 | Ba, S, O1, O2 |
| 1 | Unity | Yes | 1325 | 44.83 | 85.65 | -0.010 | Ba, S, O1 |
| 1 | Xsphere | No | 1171 | 34.15 | 75.69 | -0.129 | Ba, S, O1 |
| 1 | Xsphere | Yes | 1192 | 36.9 | 77.05 | -0.416 | Ba, S, O1 |
| 3 | Offset | No | 1206 | 33.81 | 77.96 | -0.151 | Ba, S, O1 |
| 3 | Offset | Yes | 1206 | 34.64 | 77.96 | -0.162 | Ba, S, O1 |
| 3 | Unity | No | 1291 | 44.30 | 83.45 | -0.099 | Ba, S, O1, O2, O3 |
| 3 | Unity | Yes | 1320 | 44.44 | 85.33 | -0.018 | Ba, S, O1, O2 |
| 3 | Xsphere | No | 1109 | 33.06 | 71.69 | -0.15 | Ba, S |
| 3 | Xsphere | Yes | 1163 | 36.85 | 75.18 | -0.467 | Ba, S, O1 |

* As calculated by *Sir2014* for reflections that fulfil that their intensity is above 3σ(I).

**Table S2** Merging statistics and results from *ab initio* structure solutions (*Sir2014*) of the SerialPED data collected from barite crystals on the 200kV TEM according to the different options of *partialator* in the *CrystFEL* software package. "PR ite." stands for post-refinement iterations, "Ind. Refls" for symmetrically independent reflections, and "Compl." for completeness. Overall ADP is the isotropic atomic displacement parameter obtained from the Wilson plot and given as B. Reflections up to 2 Å$^{-1}$ resolution have been considered.

| PR ite. | Partiality model | Debye Waller calculation | Ind. Refls* (#) | $R_{int}$*(%) | Compl.* (%) | Overall ADP* (Å$^2$) | Found atoms from direct methods |
|---|---|---|---|---|---|---|---|
| 0 | Offset | No  | 1272 | 18.90 | 82.22 | 0.630 | Ba, S, O1, O2 |
|   |        | Yes | 1249 | 18.79 | 80.74 | 0.598 | Ba, S, O1, O2 |
|   | Unity  | No  | 1364 | 16.48 | 88.17 | 0.673 | Ba, S, O1, O2 |
|   |        | Yes | 1364 | 16.77 | 88.17 | 0.669 | Ba, S, O1, O2 |
|   | Xsphere| No  | 1214 | 17.84 | 78.47 | 0.624 | Ba, S, O1, O2 |
|   |        | Yes | 1191 | 18.95 | 76.99 | 0.600 | Ba, S, O1, O2 |
| 1 | Offset | No  | 1256 | 18.63 | 81.19 | 0.654 | Ba, S, O1, O2 |
|   |        | Yes | 1229 | 18.82 | 79.44 | 0.684 | Ba, S, O1, O2 |
|   | Unity  | No  | 1364 | 16.5  | 88.17 | 0.673 | Ba, S, O1, O2 |
|   |        | Yes | 1364 | 16.78 | 88.17 | 0.691 | Ba, S, O1, O2 |
|   | Xsphere| No  | 1199 | 17.91 | 77.5  | 0.651 | Ba, S, O1, O2 |
|   |        | Yes | 1176 | 18.31 | 76.02 | 0.668 | Ba, S, O1, O2 |
| 3 | Offset | No  | 1209 | 18.76 | 78.15 | 0.674 | Ba, S, O1, O2, O3 |
|   |        | Yes | 1218 | 18.87 | 78.73 | 0.714 | Ba, S, O1, O2 |
|   | Unity  | No  | 1364 | 16.5  | 88.17 | 0.673 | Ba, S, O1, O2, O3 |
|   |        | Yes | 1364 | 16.76 | 88.17 | 0.744 | Ba, S, O1, O2 |
|   | Xsphere| No  | 1187 | 17.95 | 76.73 | 0.606 | Ba, S, O1, O2 |
|   |        | Yes | 1136 | 18.68 | 73.43 | 0.703 | Ba, S, O1, O2 |

* As calculated by *Sir2014* for reflections that fulfil that their intensity is above 3σ(I).

**Table S3**  Figures-of-merit for the kinematical structure refinements carried out in Jana2020 for the SerialED data collected from barite crystals on the different microscope setups. "Profile fit Int." stands for profile fit intensity extraction of the *hkl* file obtained from from *PETS2*, while "Scaling ref. Int" stands for scaling refinement intensity extraction of the *hkl* file from *diffractem*/*CrystFEL* using three post-refinement iterations. RMSD correspond to the root mean square deviation of the atom positions as calculated by *Sir2014*, and "Max. deviation" refers to the maximum distance variation of the atom positions with respect to the structure model from (Jacobsen *et al.*, 1998) as reference.

| Electron energy (keV) | **200** | | **100** |
|---|---|---|---|
| Refl. intensity extraction | **Scaling ref. Int.** | **Profile fit Int.** | **Profile fit Int.** |
| Number of reflections (#) | 2281/2717 | 267/1083 | 431/1180 |
| Reflections/Parameters (-) | 134.2 | 15.7 | 25.4 |
| GoF (%) | 43.86/40.21 | 3.74/2.44 | 4.11/2.88 |
| $R$ (%) | 52.28/55.73 | 37.59/44.27 | 34.64/39.88 |
| $R_w$ (%) | 58.34/58.39 | 42.94/47.47 | 40.40/43.21 |
| RMSD (Å) | 0.092 | 0.066 | 0.055 |
| Max. deviation (Å) | 0.407(17) | 0.30(4) | 0.23(10) |

**Table S4** Data reduction statistics and figures-of-merit for the dynamical refinements carried out in *Jana2020* for the tilt-series 3D ED data of two barite crystals.

|  | Crystal #1 | Crystal #2 |
|---|---|---|
| Angular range (°) | 120 | 100 |
| Number of patterns (#) | 121 | 101 |
| Precession angle (°) | 1.086 | 1.086 |
| Integrated Refls (#) | 37142 | 31602 |
| Merged Refls (#) | 7531 | 6395 |
| Ind. Refls* (#) | 1520 | 1216 |
| Completeness* (%) | 98.25 | 78.6 |
| $R_{int}$* (%) | 10.94 | 10.49 |
| Number of reflections (#) | 14140/16149 | 11880/13025 |
| Reflections/Parameters (-) | 91.2 | 88.0 |
| GoF (%) | 5.88/5.52 | 6.20/5.93 |
| *R* (%) | 13.48/14.02 | 13.94/14.28 |
| $R_w$ (%) | 15.30/15.34 | 15.73/15.75 |

* As calculated by *Sir2014* for reflections that fulfil that their intensity is above 3σ(I). The number of reflections, goodness of fit (*GoF*), *R* and $R_w$ parameters are calculated and reported from observed and all (obs/all) reflections up to 2 Å$^{-1}$ resolution. The criterion for observed reflections was I(**g**)>3σ(**g**). The 'Reflections/Parameters' ratio refers to the number of observed reflections over the number of refined parameters. *R* and $R_w$ are based on the square root of reflection intensities. Dynamical refinements were executed with $g_{max}$ of 2.2 Å$^{-1}$, $S_g^{max}$(matrix) of 0.01 Å$^{-1}$, $S_g^{max}$(refine) of 0.1 Å$^{-1}$, $RS_g$ of 0.66, and $N_{or}$ of 94 for crystal #1 and 98 for crystal #2.

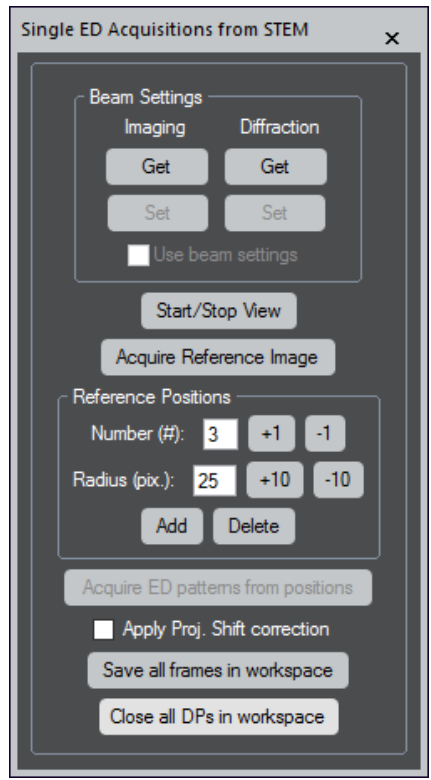

**Figure S1** Screenshot of the graphical user interface for the SerialED data collection developed in the Gatan Digital Micrograph environment.

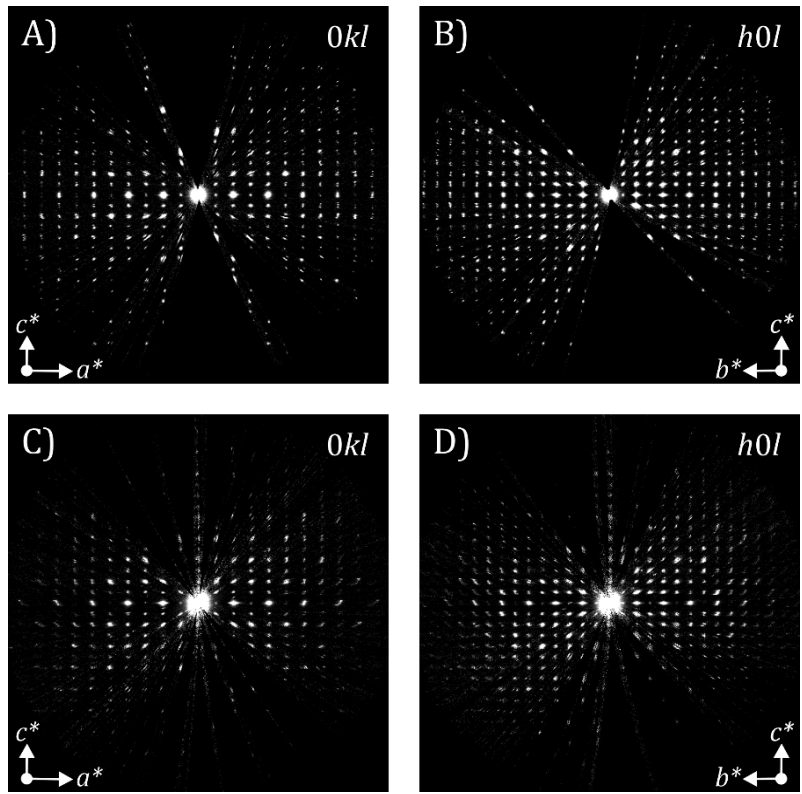

**Figure S2** 0*kl* and *h*0*l* sections of the reconstructed observable diffraction space of barite from the SerialPED data collected on the A)-B) 200kV TEM and C)-D) 100kV microscope. The low coverage along the *c\**-axis indicates the preferred orientation of the particles. Sections calculated with *PETS2* according to the found indexing.

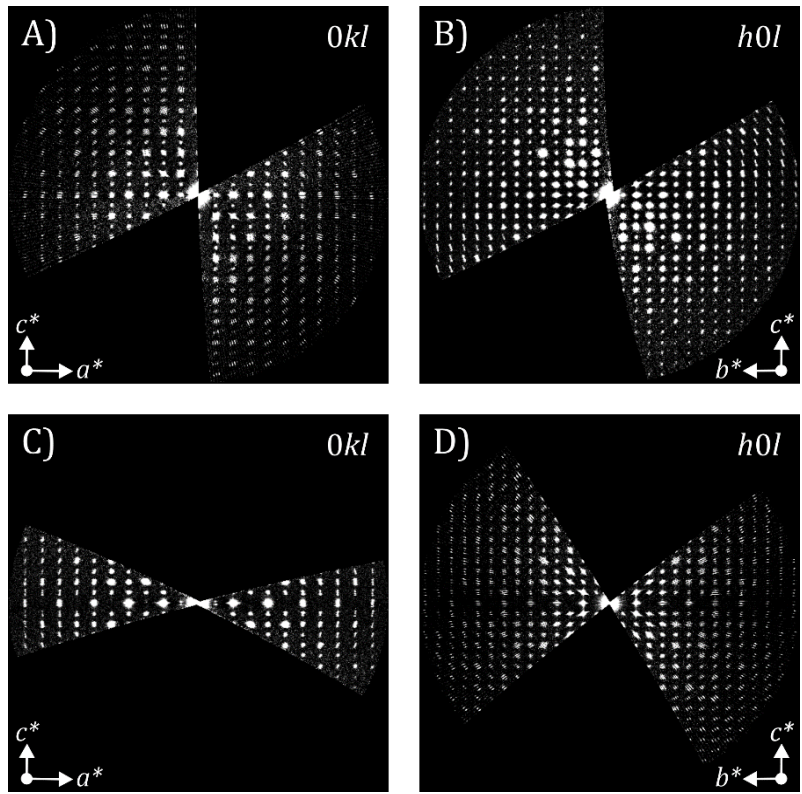

**Figure S3** *hk*0 and *h*0*l* sections of the reconstructed observable diffraction space of barite from the tilt-series 3D ED data collected of A)-B) crystal #1 and C)-D) crystal #2. Sections calculated with *PETS2* according to the found indexing.

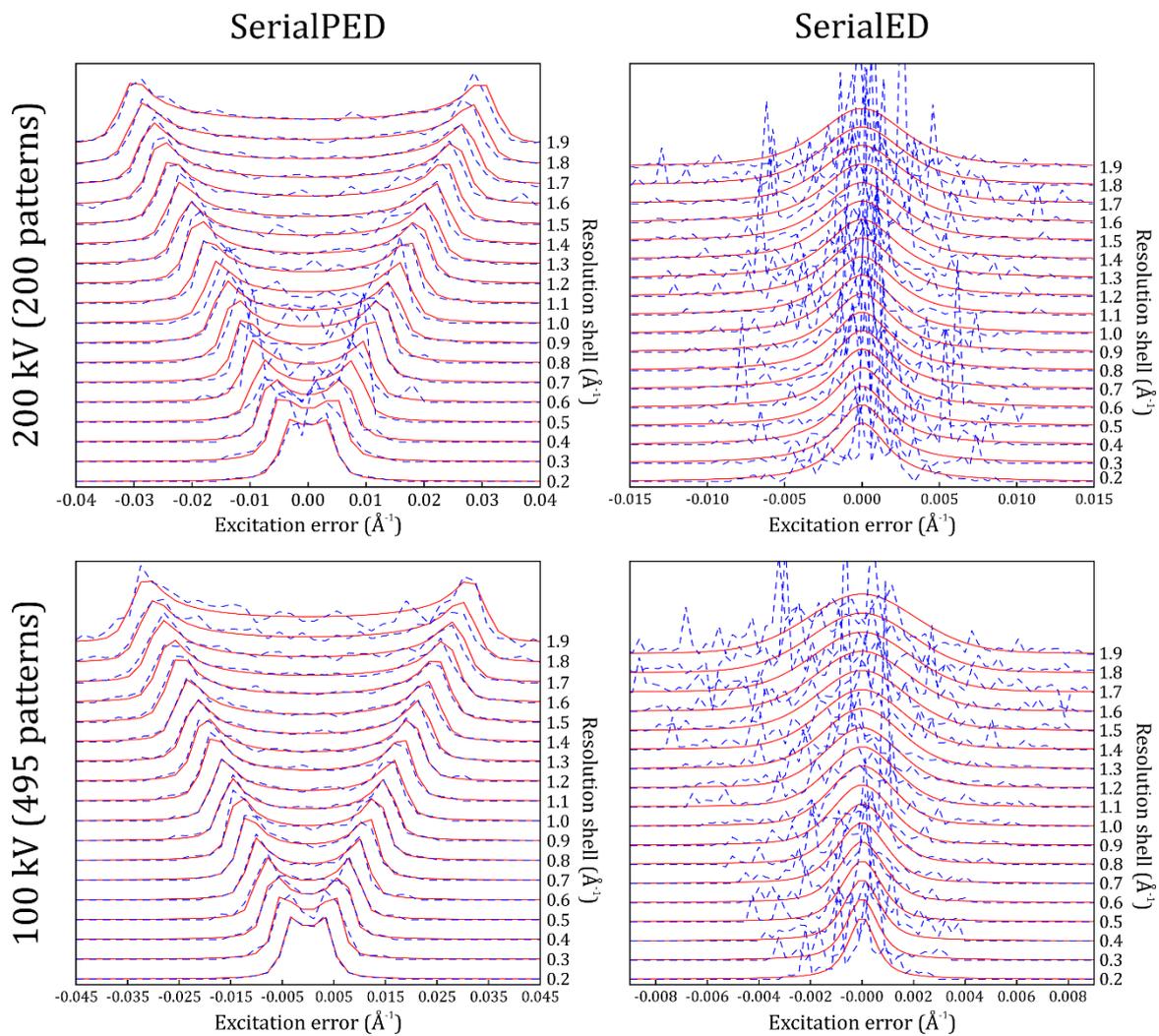

**Figure S4** Averaged rocking curves at different resolution shells for the Serial(P)ED data of barite crystals from the two different microscope setups obtained from *PETS2*. Blue-dashed curves represent the averaged experimental result and the red ones the simulated double-peaked curve that fits best.